# Sample size determination for machine learning in medical research

Wan Nor ARIFIN[#] and Najib Majdi YAACOB[#]

Biostatistics and Research Methodology Unit, School of Medical Sciences, Universiti Sains Malaysia, Kubang Kerian 16150, Kelantan, Malaysia.

[#]The authors contributed equally

**Abstract**

Machine learning (ML) methods are being increasingly used across various domains of medicine research. However, despite advancements in the use of ML in medicine, clear and definitive guidelines for determining sample sizes in medical ML research are lacking. This article proposes a method for determining sample sizes for medical research utilizing ML methods, beginning with the determination of the testing set sample size, followed with the determination of the training set and total sample sizes.

***Keywords:*** *machine learning*, *research design*, *sample size*

**Introduction**

Machine learning (ML) methods are being increasingly used in medical research, spanning various domains of medicine from oncology, orthopaedics, ophthalmology and general practice (Sirocchi et al., 2024). However, despite this advancement in medical research, currently there are no clear and definitive guidelines for determining sample sizes when using ML methods in the medical domain.

In ML, a dataset is typically divided into training and testing sets (Zhu et al., 2023). The training set is used to develop or fit the ML model (learning), while the testing set is used to evaluate the model performance (generalization) (Zhu et al., 2023; Hastie et al., 2009). The training set can be further divided into the validation set, which is used to tune the ML model prior to applying the developed model to the testing dataset (Zhu et al., 2023; Hastie et al., 2009). As opposed to the testing dataset, the validation dataset can be used repeatedly to tune the hyperparameters of the ML model. For clarity, this article divides the dataset in ML research into training and testing datasets.

Previous studies have focused sample size determinations for either the training set (Rajput et al., 2023; Riley et al., 2020; Shahinfar et al., 2020) or the testing set (Goldenholz et al., 2023; Riley et al., 2021a; Riley et al., 2021b; Archer et al., 2020) or the combined training and testing sets (Zhu et al., 2023). Goldenholz et al. (2023) proposed a model-agnostic, simulation-based method for sample size determination, while other studies provided on model-specific approaches (Riley et al., 2020; Riley et al., 2021a; Riley et al., 2021b; Archer et al., 2020). Notably, Riley et al. (2020) recommended using new data for external validation rather than splitting the dataset into training and testing sets.

To address this gap, this article proposes a method for determining sample size for medical research that utilizes ML methods that takes into account both the training and testing sets. The approach involves calculating the required sample size for the testing set using established sample size determination methods, followed by determining the sample size for the training set and total sample size. To illustrate the method, we provide a specific example focusing on the binary outcome, which is common in medical research (diseased vs non-diseased).

**Proposed Method**

In order to determine the total sample size (denoted as $n$) for a medical study utilizing ML models, the sample size can be determined for the testing set first. The sample size for the testing set is denoted as $n_{TEST}$. Depending on a predetermined training-to-testing ratio (denoted as $r_{TT}$) or proportion of testing set size ($p_{TEST}$) out of the total sample size, the sample size for the training set, denoted as $n_{TRAIN}$, and the total sample size $n$ can be determined. The steps are described as follows:

1. Define the performance metrics to be estimated using the testing set. Common metrics include diagnostic accuracy measures such as sensitivity and specificity for the binary outcome.
2. Calculate $n_{TEST}$ for estimating the selected performance metrics. For example, the sample size for estimating the sensitivity and sensitivity of the ML model in correctly classifying a medical condition can be calculated using appropriate sample size formula.
3. Calculate $n_{TRAIN}$ using $r_{TT}$.

$$n_{TRAIN} = n_{TEST} \times r_{TT}.$$

4. Calculate the total sample size $n$ using either $r_{TT}$ or $p_{TEST}$.

$$n = n_{TEST} \times (1 + r_{TT}) \text{ when using } r_{TT}, \text{ or alternatively}$$

$$n = \frac{n_{TEST}}{p_{TEST}} \text{ when using } p_{TEST}.$$

**Illustrated Example**

A (hypothetical) study aims to develop and determine the diagnostic accuracy of an ML model for the detection of COVID-19 cases based on plain chest X-rays of suspected cases of the infection. RT-PCR test results serve as the gold standard for comparison. The collected data will be split into 75% training and 25% testing sets. Based on previous studies, it can be estimated that the sensitivity and specificity will be around 85% and 75% respectively. It is estimated that 20% of the suspected cases are COVID-19 positive. The researchers aim for 5% precision for estimating these metrics with 95% confidence intervals. In total, how many suspected cases are needed? Following the steps outlined above, the calculation is as follows:

1. The performance metrics to be estimated using the testing set are sensitivity and specificity for COVID-19 status by RT-PCR = Positive, Negative.

2. For estimating the sensitivity and sensitivity of the ML model in detecting COVID-19 cases, by using a web-based calculator (https://wnarifin.github.io/ssc/sssnsp.html) (Arifin, 2024) and entering the required information provided above, the sample size for testing set is $n_{TEST} = 980$.

3. Given the data split of training:testing = 75%:25%, the $r_{TT}$ is 75/25 = 3. The sample size for $n_{TRAIN}$ is,

$$n_{TRAIN} = n_{TEST} \times r_{TT} = 980 \times 3 = 2940.$$

4. Based on the data split allocation in (3), the $p_{TEST}$ is 25/100 = 0.25. Therefore, the total sample size $n$ using $r_{TT}$ and $p_{TEST}$ is,

$$n = n_{TEST} \times (1 + r_{TT}) = 980 \times (1 + 3) = 3920 \text{ using } r_{TT}, \text{ and}$$

$$n = \frac{n_{TEST}}{p_{TEST}} = \frac{980}{0.25} = 3920 \text{ using } p_{TEST},$$

or simply $n = n_{TEST} + n_{TRAIN} = 3920$.

**Discussion and Conclusion**

This article outlines a new and practical method to determine sample sizes for ML studies in medicine, while taking into account both the training and testing sets. The increasing use of ML in medical research necessitates a general guideline in determining appropriate sample sizes for the planned studies. While previous studies have addressed specific aspects of sample size determination in this situation, a general method in determining the sample size for studies using ML is lacking. The proposed method offers a practical solution by prioritizing the testing set and subsequently determining training and total sample sizes. Given the generality of the method proposed, it can be implemented for outcomes with multiple categories or continuous variables.

**Authors' Contributions**

Conception and design: WNA, NMY

Analysis and interpretation of the data: WNA, NMY

Drafting of the article: WNA, NMY

Critical revision of the article for important intellectual content: WNA, NMY

Final approval of the article: WNA, NMY

Statistical expertise: WNA, NMY

**Correspondence**

Dr. Wan Nor Arifin

Biostatistics and Research Methodology Unit, School of Medical Sciences, Universiti Sains Malaysia, Kubang Kerian 16150, Kelantan, Malaysia.

E-mail: wnarifin@usm.my

Associate Professor Dr. Najib Majdi Yaacob

Biostatistics and Research Methodology Unit, School of Medical Sciences, Universiti Sains Malaysia, Kubang Kerian 16150, Kelantan, Malaysia.

E-mail: najibmy@usm.my

**References**

1. Sirocchi C, Bogliolo A, Montagna S. Medical-informed machine learning: integrating prior knowledge into medical decision systems. *BMC Medical Informatics and Decision Making*. 2024;24(S4). doi:10.1186/s12911-024-02582-4

2. Zhu J-J, Yang M, Ren ZJ. Machine Learning in Environmental Research: Common Pitfalls and Best Practices. *Environmental Science & Technology*. 2023;57(46):17671-17689. doi:10.1021/acs.est.3c00026

3. Hastie T, Tibshirani R, Friedman JH, Friedman JH. *The Elements of Statistical Learning: Data Mining, Inference, and Prediction*. 2nd ed. New York: Springer; 2009.

4. Rajput D, Wang W-J, Chen C-C. Evaluation of a decided sample size in machine learning applications. *BMC Bioinformatics*. 2023;24(1). doi:10.1186/s12859-023-05156-9

5. Riley RD, Ensor J, Snell KIE, et al. Calculating the sample size required for developing a clinical prediction model. *BMJ*. March 2020:m441. doi:10.1136/bmj.m441

6. Shahinfar S, Meek P, Falzon G. "How many images do I need?" Understanding how sample size per class affects deep learning model performance metrics for balanced designs in autonomous wildlife monitoring. *Ecological Informatics*. 2020;57:101085. doi:10.1016/j.ecoinf.2020.101085

7. Goldenholz DM, Sun H, Ganglberger W, Westover MB. Sample Size Analysis for Machine Learning Clinical Validation Studies. *Biomedicines*. 2023;11(3):685. doi:10.3390/biomedicines11030685

8. Riley RD, Debray TPA, Collins GS, et al. Minimum sample size for external validation of a clinical prediction model with a binary outcome. *Statistics in Medicine*. 2021;40(19):4230-4251. doi:10.1002/sim.9025

9. Riley RD, Collins GS, Ensor J, et al. Minimum sample size calculations for external validation of a clinical prediction model with a time-to-event outcome. *Statistics in Medicine*. 2021;41(7):1280-1295. doi:10.1002/sim.9275

10. Archer L, Snell KIE, Ensor J, Hudda MT, Collins GS, Riley RD. Minimum sample size for external validation of a clinical prediction model with a continuous outcome. *Statistics in Medicine*. 2020;40(1):133-146. doi:10.1002/sim.8766

11. Arifin WN. *Sample size calculator (web)* [Internet]. 2024 [cited 25 July 2024]. Available from: http://wnarifin.github.io